\documentclass{mem}
\usepackage{natbib}\usepackage{txfonts}\usepackage{balance}
\usepackage{graphicx}
\usepackage[a4paper]{hyperref}
\idline{75}{282}
\begin{document}

\title{
Three Parameters for the Horizontal-Branch Morphology in Globular Clusters.
}

   \subtitle{}

\author{
A.\,P.\,Milone\inst{1,2} 
          }

  \offprints{A.\, P.\,Milone}

\institute{
Research School of Astronomy and Astrophysics, The Australian National University, Cotter Road, Weston, ACT, 2611, Australia
\email{milone@mso.anu.edu.au}
\and
Instituto de Astrof\`\i sica de Canarias, and Department of Astrophysics University of La Laguna, E-38200 La Laguna, Tenerife, Canary Islands, Spain
}

\authorrunning{A.\, P.\, Milone}

\titlerunning{Three Parameters for the HB Morphology in GCs}

\abstract{
The horizontal branch (HB) morphology of globular clusters (GCs) is mainly described by metallicity. The fact that some clusters with almost the same metallicity exhibit different HB demonstrates that other parameters are at work.
We present results from the analysis of the CMD of 72 GCs obtained with the Advanced Camera for Survey (ACS) of the {\it Hubble Space Telescope} ({\it HST}). 
We find a significant correlation between the HB color extension and the mass of the hosting cluster, while the color distance between the HB and the red-giant branch (RGB) depends on metallicity and age. 
We suggest that age and metallicity are the main global parameters of the HB morphology in GCs, while  the HB extension is mainly due to internal helium variation, associated to multiple populations.
}
\maketitle{}

\section{Introduction}
Since the early fifties, metallicity has been considered the main parameter 
 that determines the HB morphology in GCs. 
Few years later, the evidence that some clusters with similar metallicity exhibit different HBs already suggested that at least one second parameter is required to properly characterize the HB morphology of GCs. Since then, the so called second-parameter problem has been widely investigated by many authors. 
Several candidates have been suggested as possible second parameters but a comprehensive picture is still lacking. 
We refer the reader to the papers by \cite{Catelan2009, Dotter2010, Gratton2010} for reviews. 

Recently, \cite{Dotter2010} measured the median color difference between the HB and the RGB ($\Delta$(V-I)) from ACS/{\it HST} photometry of sixty GCs and demonstrated that, after the dependence with the metallicity is accurately removed, $\Delta$(V-I) correlates with cluster age.
Also the total mass of a GC certainly plays a role on its HB morphology.
\cite{RB2006} discovered that more massive clusters tend to have HBs more extended to higher temperature. \cite{FP1993} found that the extension of the HB and the presence and extent of blue tails in particular are correlated with the cluster density and concentrations, with more concentrated or denser clusters having also bluer and more-extended HB.\\
The possibility of self-enrichment in helium as responsible of the HB shape has been investigated by several authors, as multiple stellar populations with different helium abundance can indeed explain features such us tails and multimodalities in the HBs of GCs  (e.g.\,\cite{Dantona2002, Gratton2010}).
The idea of a connection between multiple populations and HB morphology rises in the early eighties, when pioneering papers showed that the cyanogen distribution is closely connect with shape of the HB (e.g.\,\cite{Norris1981}) and is confirmed by recent studies of HB stars.\\
In this context the M\,4 represents a strong case. This GC hosts two stellar populations with different Na and O abundance that define two RGBs in the $U$ versus $U-B$ CMD. The HB of M\,4 is also bimodal and is well-populated both to the red and the blue side of the RR Lyrae gap. The bimodality in Na and O is also present among the HB. Blue-HB stars belong to the second generation and are O-poor and Na-rich, while red-HB stars are first generation (\cite{Marino2008, Marino2011}).
 Similar analysis of HB stars in other GCs show that first-generation stars populate the reddest HB segment, while second-generation HB stars have bluer colors
(e.g.\,\cite{Villanova2009, Gratton2011, Lovisi2012}).\\
In this paper we use the homogeneous photometry  from
 ACS Survey of GCs (\cite{Sarajedini2007, Dotter2011} to re-investigate the HB morphology at the light  of the new findings on multiple populations in GCs.

\section{The $L1$ and $L2$ parameters to describe the HB morphology}
\begin{figure}[t!]
\resizebox{\hsize}{!}{\includegraphics[clip=true]{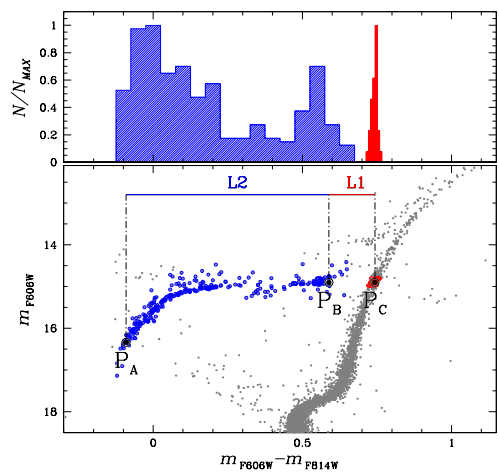}}
\caption{\footnotesize \textit{Upper panel:} Normalized histogram color distribution of stars in the HB (blue histogram) and RGB sample (red histogram) for NGC\,5904. The two sample of HB and RGB stars are colored blue and red, respectively in the lower-panel CMD, where we also show the points $P_{\rm A}$, $P_{\rm B}$, $P_{\rm C}$ 
       and the $L1$ and $L2$ segments (see text for details).}
      \label{fig1}
\end{figure}

To describe the HB, we defined two quantities: i) $L1$, which is representative of the distance between the RGB and the coolest part or the HB, and ii) $L2$ that indicates the color extension of the HB.\\
The procedure to determine $L1$ and $L2$ is illustrated in 
Fig.~\ref{fig1} for the case of NGC\,5904.
Briefly, we selected by eye a sample of HB stars, and a sample containing all RGB stars that differ by less than $\pm$0.1 F606W mag from the mean level of the HB (${\rm F606W}_{\rm HB}$, see \cite{Dotter2010}).
 Then, we have defined two points on the HB, $P_{\rm A}$ and $P_{\rm B}$, whose colors correspond to the forth and the ninety-sixth percentile of the color distribution of HB stars. The color of the third point $P_{\rm C}$ is assumed as the median color of RGB stars. $L1$ is defined as the color difference between $P_{\rm C}$ ans $P_{\rm B}$, and $L2$ as the distance between $P_{\rm B}$ and $P_{\rm A}$.

\begin{figure}[htp!]
\resizebox{\hsize}{!}{\includegraphics[clip=true]{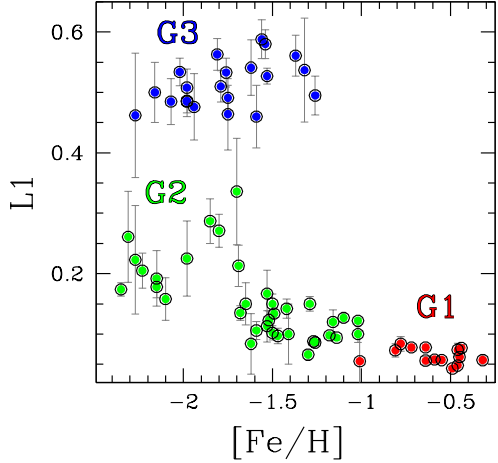}}
\caption{\footnotesize $L1$ vs.\, cluster metallicity (from Harris 1996, 2010 edition) for 72 GCs. `G1', `G2', and `G3' clusters are colored red, green, and blue, respectively. }
\label{fig2}
\end{figure}

\begin{figure}[htp!]
\resizebox{\hsize}{!}{\includegraphics[clip=true]{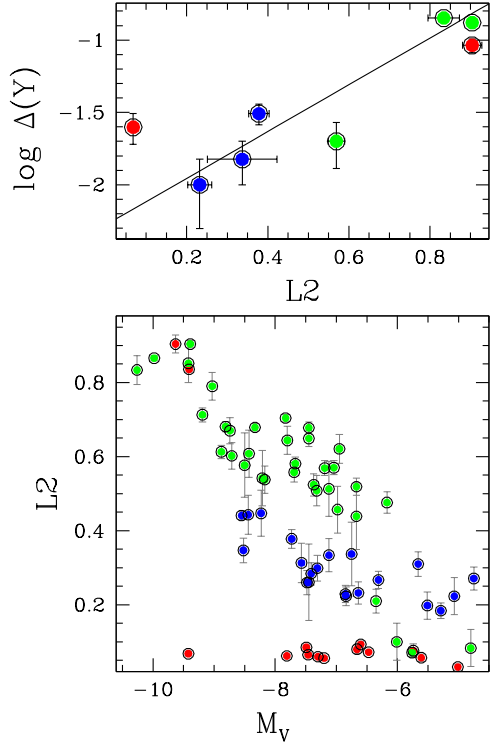}}
\caption{\footnotesize {\textit Upper panel:} logarithm of the maximum helium difference between cluster subpopulations ($\Delta$Y) as a function of $L2$.
{\textit Lower panel:} $L2$ vs.\, absolute luminosity (from Harris 1996, 2010 edition) for the GCs studied in this paper.}
\label{fig3}
\end{figure}

In the following we investigate the relation between the $L1$ and $L2$ quantities and some parameters of their host GCs.
Figure~\ref{fig2} shows $L1$ against [Fe/H].
An inspection of this plot reveals that all the metal-rich GCs have small $L1$ values and hence exhibit the red-HB. 
 At lower metallicities, there are clusters with almost the same iron abundance and  yet different $L1$ values. This reflects the well-known phenomenon that while in some GCs the red HB is absent, other clusters with almost the same metallicity host red-HB stars.\\ 
The fact that clusters populate distinct regions  in the $L1$ versus [Fe/H] plane, allows us to define three groups of GCs:
i) The first group, `G1', includes all the metal-rich GCs ([Fe/H]$>-$1.0); 
ii) the second one, `G2', is made of clusters with [Fe/H]$<-$1.0 and $L1<$0.4;
iii) the remaining GCs with $L1>$0.4 belong to the group `G3'.
Since the second-parameter phenomenon is more evident among metal-poor clusters,  we further define a fourth group (`G2+G3'), including all the GCs in the groups `G2' and `G3'. 
Both `G2' and `G3' clusters exhibit significant correlation between $L2$ and the absolute cluster luminosity (mass). This is shown in the lower panel of Fig.~\ref{fig3}, where we plot $L1$ as a function of $M_{\rm V}$. The Spearman's rank correlation coefficient is $r_{\rm G2}$=$-$0.86, and  $r_{\rm G2}$=$-$0.72 for the `G2' and `G3' sample, and $r_{\rm G1+G2}$=$-0.78$ for `G2+G3' GCs.   
There is no significant correlation between $L1$ and $M_{\rm V}$.\\
Recent papers, have shown that the CMDs of GCs 
are typically made of multiple sequences that can be followed continuously from the MS up to the RGB, and that correspond to stellar populations with different helium content (\cite{Milone201247tuc}). The maximum helium variation changes from one cluster to each other and ranges from $\Delta$Y$\sim$0.14 for the massive NGC\,2808 and $\omega$\,Cen (e.g.\, \cite{Piotto2007, King2012}) to $\Delta$Y$\sim$0.01 for NGC\,6397 (\cite{Milone20126397}). 
The upper panel of Fig.~\ref{fig3} shows that $L2$ is correlated with $\Delta$Y, in the small number of `G2' and `G3' where this measure is available. 

\begin{figure}[htp!]
\resizebox{\hsize}{!}{\includegraphics[clip=true]{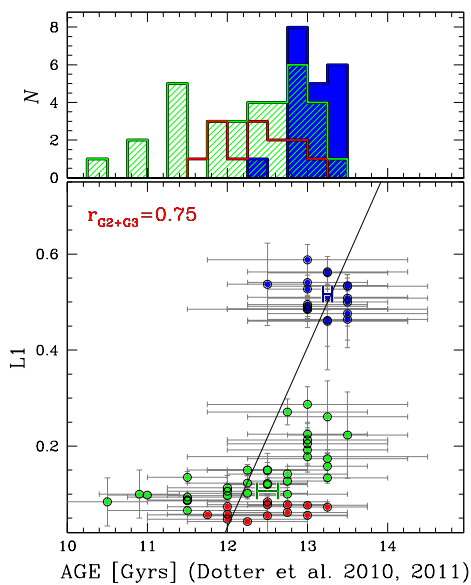}}
\caption{\footnotesize $L1$ against age (lower panel) and histogram of age distribution for the `G1' (red), `G2' (green), and `G3' clusters (blue, upper panel).
 Black line is the best-fitting straight line for the `G2+G3' sample, the Spearman coefficient $r_{\rm G2+G3}$ is also indicated.}
\label{fig4}
\end{figure}

$L1$ is plotted as a function of cluster age in the lower panel of Fig.~\ref{fig4}, while the histograms of the age distributions for `G1', `G2', and `G3' GCs are shown in the upper panel. $L1$ and age are significantly correlated for `G2+G3' clusters ($r_{\rm G1+G2}$=$0.75$), with `G3' GCs being, on average, older than   `G2' ones.
There is no significant correlation between $L2$ and age.

\section{Discussion}
\cite{Freeman1981} suggested that, apart from metallicity, at least two parameters are needed to explain the HB morphology. One of these should be
 a global parameter that varies from cluster to cluster, and the other a non-global parameter that varies within the cluster.\\
Our analysis reveals that the color distance between the RGB and the coolest part of the HB, $L1$, depends on cluster age and metallicity, while the HB extension, $L2$, correlates with the cluster luminosity (and hence the mass).
Recent works on multiple stellar populations in GCs show that  massive clusters exhibit, on average, larger internal helium variations, $\Delta$Y, than small-mass GCs. As expected, we found that $\Delta$Y is positively correlated with $L2$, even if this analysis is limited to a small number of clusters (see also \cite{Dantona2002, Dantona2008, Gratton2010} for discussion on the connection between helium and HB morphology).\\
These results suggest that age and metallicity are the main global parameters of the HB morphology of GCs, while internal star-to-star helium variation, associated to the presence of multiple populations, are the main responsible of the HB extension.

\begin{acknowledgements}
Results of this paper come from the work of several persons. I thank
J.\,Anderson, A.\,Aparicio, L.\,R.\,Bedin, A.\,Dotter,  A.\,F. \, Marino, M.\,Monelli, G.\,Piotto, A.\,Recio Blanco, A.\,Sarajedini and the ACS GCs treasury group.  
\end{acknowledgements}

\bibliographystyle{aa}

\end{document}